\documentclass[%
reprint,
 amsmath,amssymb,mathtools,
 aps,
prstab,
]{revtex4-2}
\usepackage[version=3]{mhchem}

\usepackage{graphicx}% Include figure files
\usepackage{dcolumn}% Align table columns on decimal point
\usepackage{bm}% bold math
\usepackage{amsmath,booktabs,siunitx,physics}
\begin{document}
%\sffamily\mgfamily

\preprint{FFA/007-SHE}

\title{Design of a Storage Ring based on  \\
  a Fixed Field Alternating Gradient Configuration with  an Internal Target for \\
  Heavy-Ion Beams with Stochastic Charge State Conversions}% Force line breaks with \\
%\thanks{A footnote to the article title}%

\author{Yoshihiro Ishi}
\email{ishi.yoshihiro.4m@kyoto-u.ac.jp}
\author{Tomonori Uesugi}
\author{Yoshiharu Mori}%
\affiliation{%
 Institute for Integrated Radiation and Nuclear Science, Kyoto University, Kumatori, Osaka 590-0494, Japan}

\author{Katsuhisa Nishio}
% \altaffiliation[Also at ]{Physics Department, XYZ University.}%Lines break automatically or can be forced with \\
% \email{Second.Author@institution.edu}
\affiliation{%
  Advanced Science Research Center, Japan Atomic Energy Agency, Tokai, Ibaraki 319–1195, Japan
}

\date{\today}% It is always \today, today,
             %  but any date may be explicitly specified

\begin{abstract}
  In the general use of a heavy-ion accelerator, an accelerated beam impinged on a target is spoiled into a beam dump.
  To make more efficient use of the beam,
  recycling of the beam passed through
  the target is proposed
  in the framework of
  the so-called energy recovery internal target (ERIT). In the ERIT system, the target is irradiated inside the circulating
  beam by recovering the energy lost in the target using
  rf cavities.
  So far, such a system has been realized only for proton beams. 
  Here, the ERIT system for heavy-ion beam 
  is demonstrated for the first time. A challenging issue is the circulation of all ions with different atomic charge.
  An ion has a probability of equilibrated charge state distribution after passing through the target,
  independent of the initial charge state. 
  This phenomenon of stochastic charge state conversion (SCSC) causes rapid beam-emittance growth.
  To solve this problem,
  we developed a method to match the closed orbits and betatron functions of the beams in different
  charge states at the target location in a ring based on a scaling fixed field alternating gradient (FFA) lattice structure.
  We present the design of such an FFA ring and show, through full 6D beam tracking simulations, that transverse emittance growth  
  can be effectively suppressed even in the presence of SCSC.
\end{abstract}

\maketitle

%%%%%%%%%%%%%%%%%%%%%%%%%%%%%%%%%%%%%%%%%%%%%%%%%%%%
\section{\label{sec:intro}introduction}
%%%%%%%%%%%%%%%%%%%%%%%%%%%%%%%%%%%%%%%%%%%%%%%%%%%%

Heavy-ion accelerators have been the subject of research and development for decades because of their versatile applications~\cite{ikegami, steck, franzke, kandler, bruno, litvinov, ghazaly, olsen, review_of_HISR, cosmic_atomic_plasmas}, from fundamental physics research programs, such as superheavy element production, to medical isotope production~\cite{oganesian_cyclo,at211}.
The production rate of the desired nuclides/isotopes depends on the number of beam particles passing through the target
at a unit time interval. In the traditional method, accelerated heavy ions  are spoiled into a beam dump after traversing the target. The only way to increase the reaction rate is to develop a high-intensity ion source and dedicated accelerator system.

When one tries to create new elements $Z$=120 in the reaction of \ce{^{50}Ti} beam and \ce{^{249}Cf} target, for example, the predicted cross-section deviates by two orders of magnitude from a femtobarn (fb) to \SI{100}{fb} between the theories~\cite{gates}. The 1-fb experiment requires approximately 10 years using a \SI{1}{p\micro A} beam and a \SI{500}{\micro g/cm^2} target, which is
a typical condition of the current setup for superheavy element research.

To significantly increase a reaction rate and reduce the beam time, a method to repeatedly use the beam which passed through the target can be considered.
Such an idea is called the energy recovery internal target(ERIT)~\cite{ERIT} system using a storage ring,
in which the internal target is irradiated by a circulating beam.
The energy lost in
the target is recovered by the rf cavities.
So far, this approach has been adopted only for proton beam.
To adopt this approach for heavy-ion beams, for example, in the field of superheavy element research at
a beam energy of approximately 5--6\SI{}{MeV/u}, all the beams with different atomic charge states must be circulated
with a proper beam area/size at the target.

If the ion in the beam has a small nuclear charge $Z$ and sufficient energy, the electrons of the atoms are fully stripped.
However, if the beam has a large $Z$ and a relatively low energy insufficient to fully strip the electrons,
the atomic charge states $q$ have a equilibrated distribution the Gaussian-like shape, irrespective of initial charge state,
after passing through the target.
This stochastic charge state conversion (SCSC) at the target poses a challenge for heavy-ion ERIT systems.
In this article, we show that the fixed-field alternating-gradient (FFA) configuration can be used to realize an
ERIT system for heavy-ion beams, even in the presence of SCSC.

\begin{figure}[htb]
\centering
    \includegraphics[clip,width=8cm]{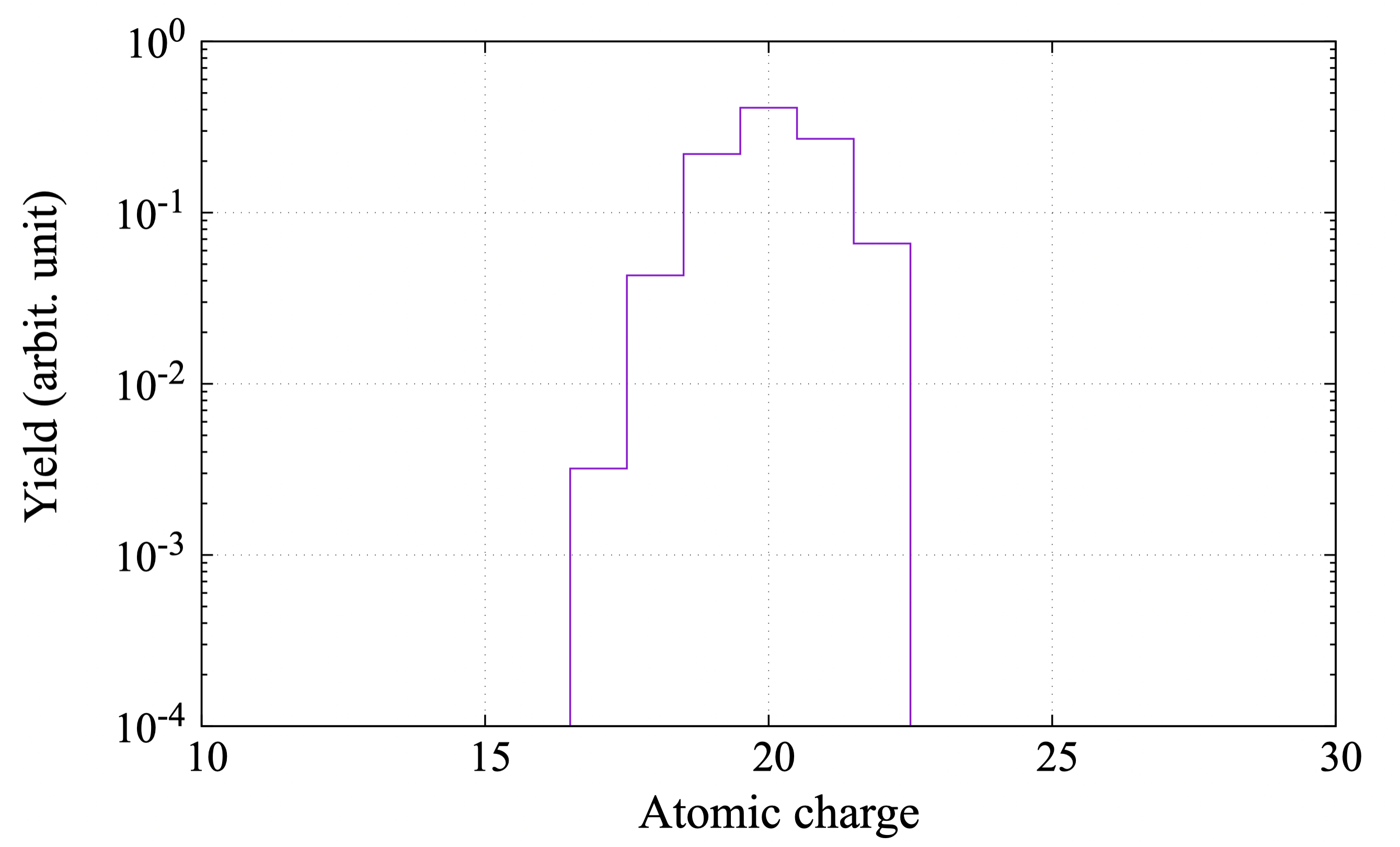}
    \caption{Equilibrium atomic charge state distribution calculated by using Shima's formula 
      for the case of \ce{^{50}Ti} beam  of \SI{5.5}{MeV/u}}.
\label{fig:shima_dist}
\end{figure}

Figure~\ref{fig:shima_dist} shows the atomic charge state distribution for \ce{^{50}Ti}
  beam of 5.5 MeV/u,
  one of the candidate beam to create
  superheavy elements, calculated using the Shima formula~\cite{shima}. It has the maximum fraction at $q = 20^+$
  and ranges from $17^+$ to $22^+$.
  This corresponds to a momentum spread of $\Delta p/p$ from $-15\%$ to $+10\%$ around $q$=$20^+$ , which leads to a betatron tune
  (hereafter simply referred to as "tune")
  spread due to chromaticity.
In the conventional separated-function lattices, nonlinear elements such as sextupole magnets are used to correct chromaticity.
However, it is difficult to compensate for a wide momentum spread exceeding $\pm10\%$, as in the present case.
Even when correction is effective, the dynamic aperture can be significantly reduced due to strong nonlinearities.
In contrast, scaling FFA rings inherently maintain constant tunes regardless of beam momentum,
since the chromaticity is zero~\cite{symon}.
Another notable advantage of scaling FFA rings is their large dynamic aperture, which is preserved even under zero-chromaticity conditions.

\begin{figure}[htb]
\centering
    \includegraphics[clip,width=8cm]{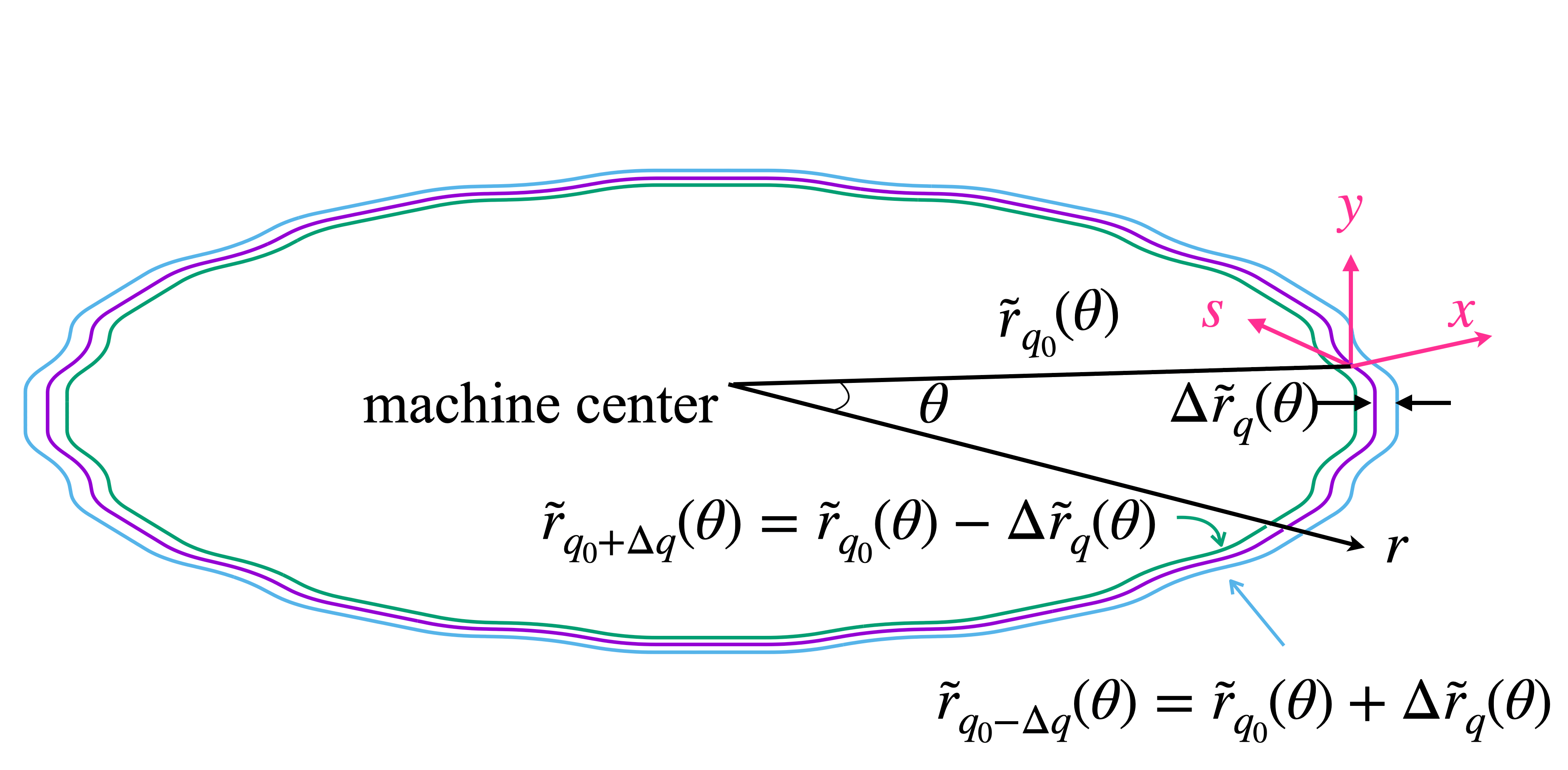}
\caption{Definition of the coordinate system and closed orbits.The orbit is drawn for the FFA having 12 FDF sectors.}
\label{fig:varidef}
\end{figure}

  In general,
the magnetic field distribution in the median plane of a scaling FFA ring is~\cite{symon}
\begin{align}
  \label{eq:scaling_ffa}
  B(r)\propto r^{k_0} ,
\end{align}
where $k_0$ is a magnetic field index that is constant in the horizontal $r$ and azimuthal $\theta$ directions.
The definitions of the variables used in this study are summarized in Table~\ref{tab:varidef}.
As shown in Fig.~\ref{fig:varidef}, the closed orbit for charge state $q$ is denoted by $\tilde{r}_{q}(\theta)$.
The charge difference relative to $q_0$ is denoted by $\Delta q$,
where $q_0$ is the nominal charge--{\it i.e.}, the most probable charge (20$^+$ in the present case).
The orbit separation between $q_0$ and $q_0 \pm \Delta q$ is represented by $\Delta\tilde{r}_q(\theta)$.
\begin{table*}[htb]
  \setlength\tabcolsep{5.5pt}
  \centering
  \caption{Definition of the variables.}
  \label{tab:varidef}
  \begin{tabular}{l l }
  \hline
  \hline
  $r$          & Radial coordinate in the global cylindrical coordinate system  \\
  $\tilde{r}(\theta)$  & Radius of the closed beam orbit as a function of $\theta$ \\
  $\tilde{r}_{q}(\theta)$  & Radius of the closed beam orbit in the charge state $q$ \\
  $\tilde{r}_{q_0}(\theta)$  & Radius of the closed beam orbit in the nominal charge state \\
  $\Delta\tilde{r}_q(\theta)$  & Difference between  $\tilde{r}_{q_0-\Delta q}(\theta)$ and $\tilde{r}_{q_0}(\theta)$ \\
  $\theta$     & Azimuthal coordinate in the global cylindrical coordinate system \\
  $\theta_F$   & $\theta$ evaluated at the center of the F magnet \\
  $\theta_D$   & $\theta$ evaluated at the center of the D magnet \\
  $\theta_T$   &  $\theta$ evaluated at the target position \\
  \hline
  $x$  & Radial (or horizontal) coordinate in the local Frenet-Serret coordinate system on the closed orbit \\
  $y$  & Binormal (or vertical) coordinate in the local Frenet-Serret coordinate system on the closed orbit \\
  $s$  & Tangential (or along the beam axis) coordinate in the local Frenet-Serret coordinate system on the closed orbit \\
  $x_{cod}$ & Horizontal closed orbit distortion (closed orbit displacement in $x$) \\
  2$J_x$ & Horizontal betatron amplitude: $2J_x =  \gamma_x x^2 + 2\alpha_x xx^\prime + \beta_x x^{\prime 2}$, $\alpha_x = -\frac{1}{2}\frac{d\beta_x}{ds}, \gamma_x= \frac{1+\alpha_x^2}{\beta_x}$ \\
  2$J_y$ & Vertical   betatron amplitude:  $2J_y =  \gamma_y y^2 + 2\alpha_y yy^\prime + \beta_y y^{\prime 2}$, $\alpha_y = -\frac{1}{2}\frac{d\beta_y}{ds}, \gamma_y= \frac{1+\alpha_y^2}{\beta_y}$ \\
  \hline
  \hline
  \end{tabular}
\end{table*}
From Eq.~(\ref{eq:scaling_ffa}), the relationship of the charge difference and
the closed orbit separation is given by,
\begin{align}
  \label{eq:momentum_compaction}
  \Delta \tilde{r}_q(\theta)= 
  - \frac{\Delta q}{q_0}\frac{\tilde{r}_{q_0}(\theta)}{k_0+1},
\end{align}
where  $\tilde{r}_{q_0}(\theta)/(k_0+1)$ corresponds to the dispersion function.

\begin{figure}[htb]
    \centering
    \includegraphics[width=8cm]{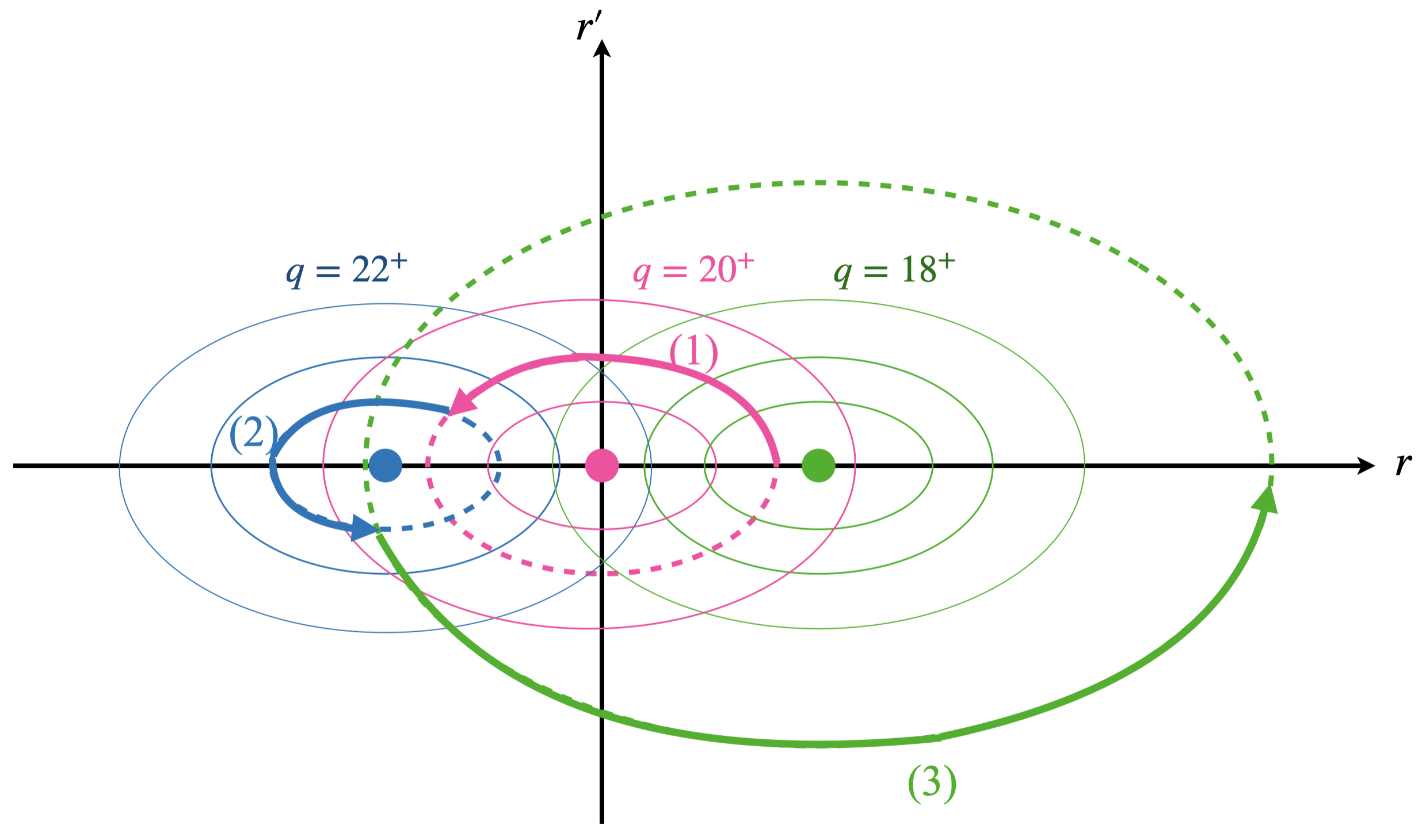}
    \caption{Example of the change of betatron amplitude  resulting from charge state changes:
      $q = 20^+ \rightarrow 22^+ \rightarrow 18^+$, corresponding to the amplitude shift (1) $\rightarrow$ (2) $\rightarrow$ (3).}
    \label{fig:emit_dipole}
\end{figure}

As shown in Fig.~\ref{fig:emit_dipole}, the closed orbit point--i.e., the center of the Courant-–Snyder ellipse--
in transverse phase space at the target fluctuates when the SCSC occurs. As a result,
the betatron amplitude changes accordingly.
Although the betatron amplitude increases or decreases depending on the betatron phase at the target,  
it tends to grow through long-term stochastic collision processes (see Fig.~\ref{fig:sherit}).  
To suppress this growth, the closed orbits for different charge states must satisfy the following achromatic conditions:
\begin{align}
  \label{eq:achromat}
  \tilde{r}_{q_0}(\theta_T) &= \tilde{r}_{q_i}(\theta_T),  \\
  \label{eq:achromat2}
  \tilde{r}^\prime_{q_0}(\theta_T) &= \tilde{r}^\prime_{q_i}(\theta_T),
\end{align}
where $\theta_T$ is the azimuthal angle at the target location, $q_i$ is any possible charge state,  
and $\tilde{r}^\prime$ denotes the derivative of $\tilde{r}$ with respect to the longitudinal coordinate~$s$.

Achromatic lattices using FFA-type magnets have been studied, and several types of dispersion suppressors
have been proposed~\cite{machida,adam,ishi_ipac23,JB}.
The lattice introduced in Ref.\cite{machida} provides a dispersion suppressor in the straight beam transport line,
which is not suitable for our purpose.
According to studies in Ref.\cite{adam,ishi_ipac23,JB}, the horizontal phase advance in the arc must be an
integer multiple of $2\pi$ to suppress the dispersion to zero.
This implies that a tune-adjusting straight section is required to avoid an integer horizontal tune.
However, such straight sections break the condition of zero chromaticity.
Furthermore, this approach is difficult to apply to storage rings, which require both zero chromaticity
and a horizontal tune sufficiently far from integer values.
Therefore, we propose an alternative method to satisfy the conditions in Eqs.~(\ref{eq:achromat}) and (\ref{eq:achromat2}).
  
The remainder of this paper is organized as follows: Section~\ref{sec:method} describes the lattice construction
method; Section~\ref{sec:sim} presents the results of the beam simulations; Section~\ref{sec:disc} discusses beam stability associated with the use of modulated $k$ and the space-charge effects; and Section~\ref{sec:concl} provides the conclusion.

%%%%%%%%%%%%%%%%%%%%%%%%%%%%%%%%%%%%%%%%%%%%%%%%%%%%
\section{Method of the lattice construction}
\label{sec:method}
%%%%%%%%%%%%%%%%%%%%%%%%%%%%%%%%%%%%%%%%%%%%%%%%%%%%
In this section, we describe the design procedure of a heavy-ion storage ring intended for the production of superheavy elements.
The proposed “test ring” is based on a 12-fold symmetric radial scaling FFA and comprises 12 FDF (Focus–Defocus–Focus) triplets.
This type of lattice has been previously adopted and experimentally tested in the ERIT system at the Institute for Integrated Radiation and Nuclear Science, Kyoto University (KURNS)\cite{ERIT_okabe, MERIT}.
The main design parameters of the test ring are summarized in Table\ref{tab:main_parameter}.

  The magnetic field distribution in the test ring takes a more specific form of Eq.(\ref{eq:scaling_ffa}), given by
\begin{align}
\label{eq:scaling_ffa2}
B(r) = B_0 \left( \frac{r}{r_0} \right)^{k_0},
\end{align}
where $B_0$ and $r_0$ are the reference magnetic field and reference radius, respectively,
specified for the F and D magnets as listed in Table\ref{tab:main_parameter}.
\begin{table}[htb]
  \setlength\tabcolsep{5.5pt}
  \centering
  \caption{Basic parameters of the test ring.}
  \label{tab:main_parameter}
  \begin{tabular}{lcc}
  \hline
  \hline
    Beam species &  \multicolumn{2}{c}{\ce{^{50}Ti^{q+}} (q = 17--22)} \\ 
    Beam Energy & \multicolumn{2}{c}{\SI{5.5}{MeV/u}} \\
    Lattice &  \multicolumn{2}{c} {Radial sector FDF triplet} \\
    Symmetry & \multicolumn{2}{c} {12} \\
  \hline
    Field index & $k_0$ &  2.2 \\
    Betatron tune & $(\nu_x,\nu_y)$ & (1.91, 3.08) \\
  \hline
    $B_0(=B(r_0))$ & F magnet & \SI{0.725}{T}  \\
          & D magnet & \SI{-0.4723}{T} \\
    reference radius $r_0$ & F magnet & \SI{5.97}{m}  ($=\tilde{r}_{q_0}(\theta_F$))$^{\rm i}$ \\
          & D magnet & \SI{5.82}{m}  ($=\tilde{r}_{q_0}(\theta_D$)) \\
    opening angle & long straight  & \SI{10.5}{\degree} \\
                  & short straight &  \SI{1.7}{\degree} \\
                  & F magnet       &  \SI{5.12}{\degree} \\
                  & D magnet       & \SI{6.86}{\degree} \\
  \hline
  \hline
  \end{tabular}
  \\ i~Radius of the closed orbit at the center of F magnet
\end{table}

%-----------------------------
\subsection{Matching of closed orbit and betatron functions }
\label{subsec:cod_matching}
\begin{figure}[htb]
\centering
    \includegraphics[width=8cm]{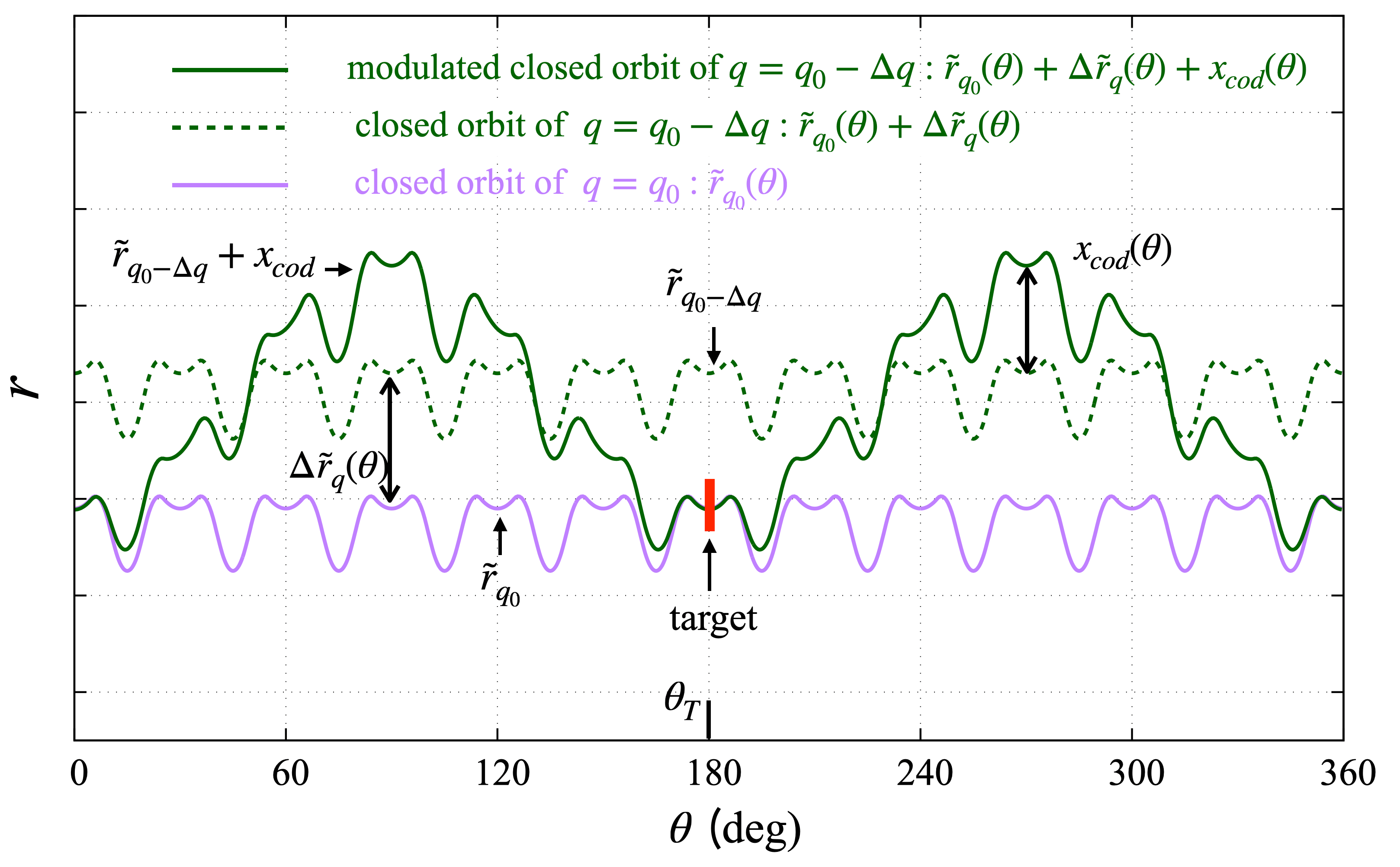}
    \caption{
      Closed orbit matching. The purple line indicates the closed orbit of the beam in the nominal charge state $q_0$.
      The dashed/solid green lines indicate the unmodulated/modulated closed orbits of the charge $q_0 - \Delta q$ beam.}
\label{fig:cod_matching}
\end{figure}

%-----------------------------
To fulfill the condition of Eq.~(\ref{eq:achromat}), we intentionally use a closed orbit distortion~(COD) $x_{cod}(\theta)$
as shown in Fig.~\ref{fig:cod_matching} so that
\begin{align}
  \label{eq:cod_matching_cond}
  \Delta \tilde{r}_q(\theta_T) + x_{cod}(\theta_T) &= 0.
\end{align}
The condition of Eq.~(\ref{eq:achromat2}) is automatically satisfied if the target is located at the point where
the mirror symmetry of the lattice is maintained.

\begin{figure}[h]
\centering
    \includegraphics[width=8cm]{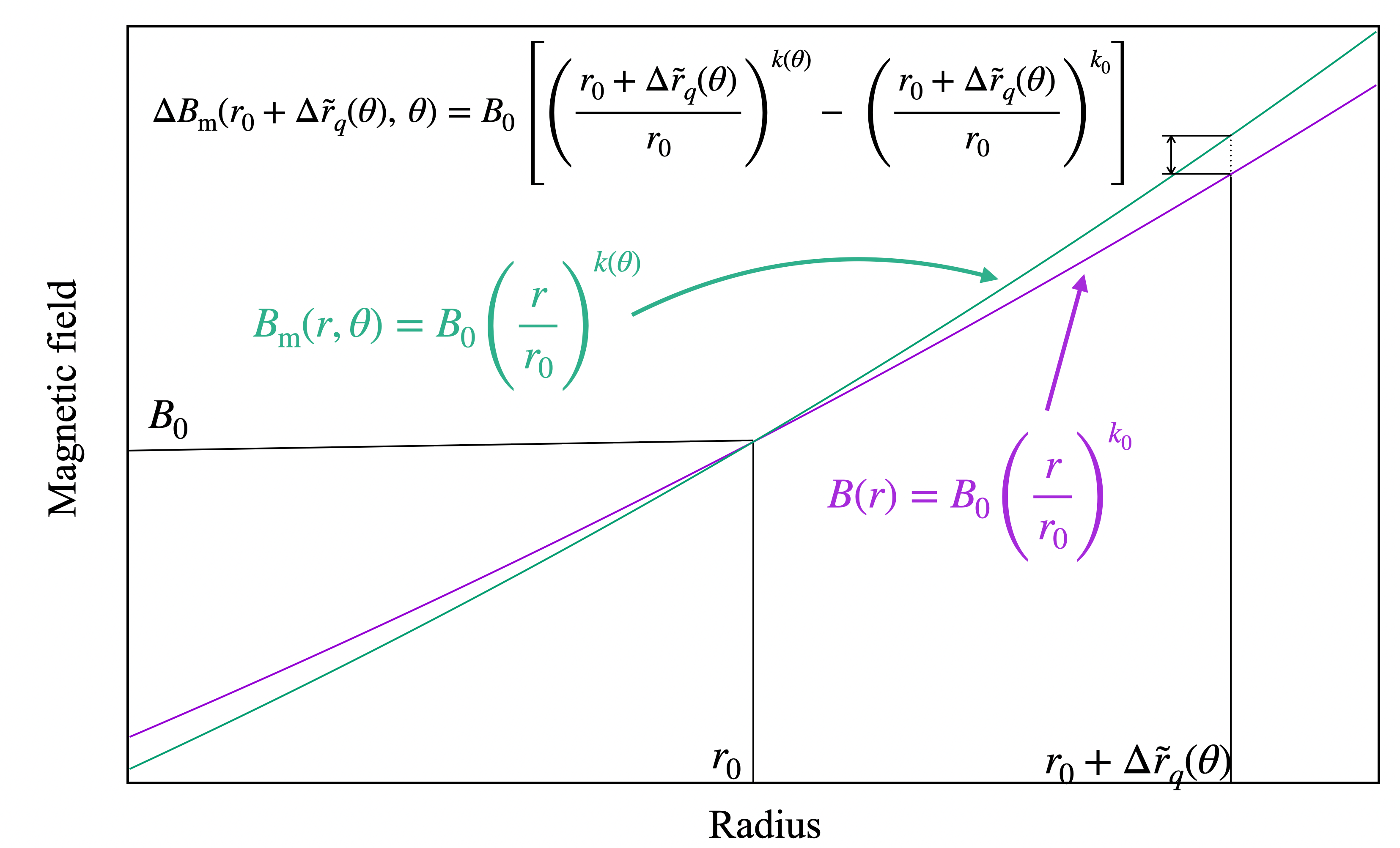}
    \caption{The $r$-dependence of the magnetic field at the center of the F magnet as an example.
      The purple and green lines describe the magnetic field as functions of $r$,
      obtained using constant~$k$ and modulated~$k$, respectively.}
\label{fig:bmod}
\end{figure}

Now, we consider the case in which the field index is not strictly constant with $k_0$
but slightly modulated in $\theta$ direction as
\begin{align}
\label{eq:k_mod}
k(\theta) = k_0 + \Delta k(\theta),
\end{align}
which modulates the magnetic field as
\begin{align}
  \label{eq:b_w_mod}
  B_{\text{m}}(r,\theta)&=B_0\left(\frac{r}{r_0}\right)^{k(\theta)},% \notag \\
\end{align}
where $B_0$ is the same definition as in Eq.~(\ref{eq:scaling_ffa2}).
For example, $r$-dependence of magnetic field at the center of F magnet ($\theta=\theta_F$) is shown in
Fig.~\ref{fig:bmod}, where
the purple and green lines describe $B(r)$ with a constant $k(=k_0)$ and $B_{\text{m}}(r,\theta)$ with a
modulated $k(=k(\theta))$, respectively. 
Along the closed orbit of charge $q_0-\Delta q$ beam $r(\theta)=\tilde{r}_{q_0}(\theta)+\Delta\tilde{r}_q(\theta)$
the magnetic field modulation which should be introduced is given by 
\begin{align}
  \label{eq:dipole_mod}
  \Delta &B_{\rm m}(r_0+\Delta \tilde{r}_q(\theta),\theta) \notag \\
  & =  B_0
  \left[ \left(1+\frac{\Delta\tilde{r}(\theta)}{r_0}\right)^{k_0+\Delta k(\theta)}
    - \left(1+\frac{\Delta\tilde{r}(\theta)}{r_0}\right)^{k_0} \right]  \notag \\
  & =  B_0 \left(\frac{\Delta\tilde{r}(\theta)}{r_0}\right)\Delta k(\theta)
  + \mathcal{O}\left[\left(\frac{\Delta\tilde{r}(\theta)}{r_0}\right)^2\right] \notag \\
  &    \sim B_0\frac{\Delta k(\theta)}{k_0+1}\frac{\Delta q}{q_0} , 
\end{align}
where we used the general characteristics of the scaling FFA
described in Eq.~(\ref{eq:momentum_compaction}).

In general, if there is an extra dipole field distributed
as $\mathcal{E}(\theta)$
along the closed orbit,
a COD will emerge and the displacement from the original closed orbit $x_{cod}(\theta)$ is represented by the
Fourier series as~\cite{COURANT_SNYDER}
\begin{align}
        \label{eq:cod}
        x_{cod}(\theta) & = \sqrt{\beta_x(\theta)}\sum_{n=-\infty}^{\infty}\frac{\nu_x^2}{\nu_x^2-n^2} \tilde{f}_n e^{in\phi_x}, 
\end{align}
where $\nu_x$ is the horizontal tune, $\beta_x$ is the horizontal betatron function and
$\phi_x$ is a normalized phase advance
\begin{align}
  \phi_x(s) = \frac{1}{\nu_x}\int^s_0\frac{ds^\prime}{\beta_x(s^\prime)},
\end{align}
which advances $2\pi$ per turn in the ring.
We use a smooth focusing approximation of $\phi_x\sim\theta$ to simplify the design optimization process.
In Eq.~(\ref{eq:cod}), $\tilde{f}_n$ is the $n$-th Fourier component, defined as
\begin{align}
\label{eq:fn}
\tilde{f}_n & = \frac{1}{2\pi B\rho}\oint e^{-in\theta}
\beta^{3/2}(\theta)\mathcal{E}(\theta) d\theta,
\end{align}
where $B\rho$ is the magnetic rigidity of the beam with charge state $q_0 - \Delta q$.
Because this modulation breaks the orbit similarity, beams with different charge states have different COD patterns. We attempt to make these closed orbits converge into the same position in the $r$-direction at the target location $\theta_T$, while the orbit of the beam with the nominal charge $\tilde{r}_{q_0}(\theta)$ does not have any modulation.
By substituting $\Delta B_{\mathrm{m}}(\tilde{r}_{q_0}(\theta) + \Delta \tilde{r}_q(\theta), \theta)$ into the perturbation
term $\mathcal{E(\theta)}$, Eq.~(\ref{eq:fn}) is rewritten as:
\begin{align}
        \label{eq:fn2}
     \tilde{f}_n    = \frac{1}{2\pi B\rho}\frac{\Delta q}{q_0} \frac{B_0}{k_0+1}\oint e^{-in\theta}
  \beta^{3/2}(\theta) k(\theta) d\theta.
\end{align}
Using Eq.~(\ref{eq:dipole_mod}), (\ref{eq:cod}) and (\ref{eq:fn2}) the convergence condition Eq.~(\ref{eq:cod_matching_cond}) becomes
\begin{align}
  \label{eq:convcond}
  \frac{\Delta q}{q_0(k_0+1)}\left(\tilde{r}_{q_0} + \frac{\sqrt{\beta(\theta_T)}}{2\pi B\rho}\sum_n\frac{e^{in\theta_T}\Phi_n[\Delta k]}{\nu_x^2-n^2}  \right) =0,
\end{align}
where we define
\begin{align}
\Phi_n[\Delta k] := B_0 \oint e^{-in\theta}
  \beta_x^{3/2}(\theta) \Delta k(\theta)d\theta.
\end{align}
Because the condition for $\Phi_n[\Delta k(\theta)]$ in Eq.~(\ref{eq:convcond}) is independent of $\Delta q$, the closed orbit in any charge state at the target location can converge to a point if we appropriately set the modulation $\Delta k(\theta)$.

To prevent SCSC-induced stochastic fluctuations of the Courant–Snyder ellipse and the resulting growth in betatron amplitude,
it is essential to match not only the center of the ellipse (i.e., the closed orbit), but also its shape.
This requires that the Courant–Snyder parameters $\beta_x$, $\beta_y$, $\alpha_x$, and $\alpha_y$ be matched at the target location.
Just as modulation of the dipole component leads to a COD,
modulation of the quadrupole component affects the betatron functions.
The resulting modulation of $\beta_x$ is given by~\cite{COURANT_SNYDER}:
\begin{align}
\label{eq:beta_mod}
\frac{\Delta\beta_x}{\beta_x} = -\frac{\nu_x}{2} \sum_{p_x=-\infty}^{+\infty} \frac{\tilde{g}_{p_x} e^{ip_x \phi_x}}{\nu_x^2 - \left(\frac{p_x}{2}\right)^2},
\end{align}
where $\tilde{g}{p_x}$ is the Fourier coefficient corresponding to harmonic number $p_x$, given by
\begin{align}
\tilde{g}_{p_x} = \frac{\nu_x}{2\pi} \int_0^{2\pi} \Delta K_x(\phi_x) \beta_x^2(\phi_x) e^{-ip_x \phi_x}  d\phi_x,
\end{align}
and $\Delta K_x$ denotes the modulation of the quadrupole component of the magnetic field.
The same expression applies to the vertical beta function $\beta_y$ by replacing the subscript $x$ with $y$.
Since modulation of the field index $k$ also introduces quadrupole modulation,
it can be utilized to control and match the betatron functions.
%-----------------------------
\subsection{Optimization of the $k$ modulation}
\label{subsec:opt_k_mod}
%-----------------------------

\begin{figure*}[htb]
\centering
    \includegraphics[width=16cm,clip]{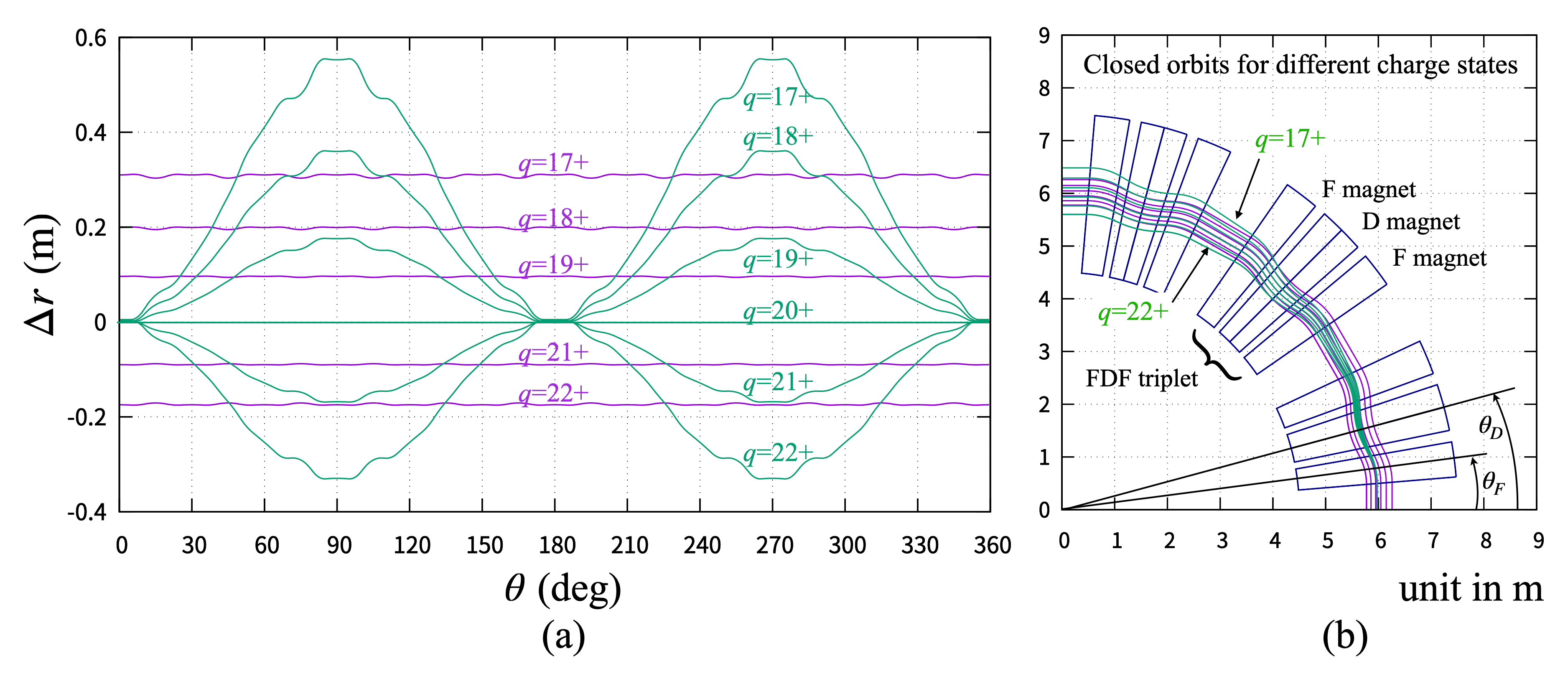}
    \caption{
      The plots in (a) show the deviation of the closed orbits of various charge states from 17$+$ to 22$+$ concerning the closed orbits of the 20$+$ charge state, with purple and green lines for the constant~$k$ and modulated~$k$ rings, respectively. In (b), the identical closed orbits in (a) are plotted as footprints.
}
\label{fig:orbits}
\end{figure*}
\begin{figure*}[htb]
\centering
    \includegraphics[width=16cm,clip]{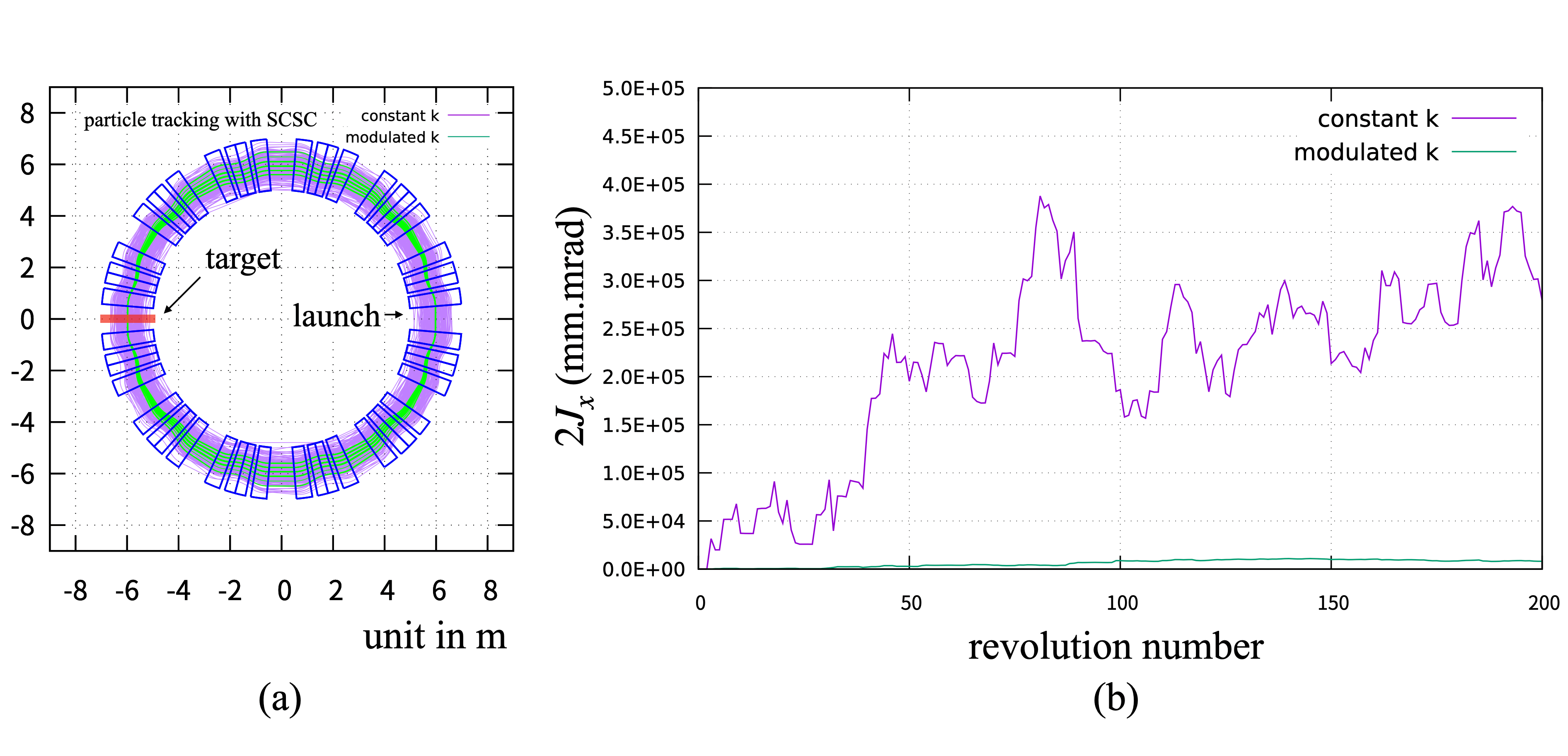}
    \caption{
The plots in (a) show the footprints of single particle tracking during 200 turns, considering the SCSC with purple and green lines for the constant~$k$ and modulated~$k$ rings, respectively. The plots in (b) show the horizontal betatron oscillation amplitude $2J_x$ as a function of the revolution number.      
    }
\label{fig:sherit}
\end{figure*}

Using the harmonic numbers $n$, $p_x$, and $p_y$ , we apply the following modulation to $k(\theta)$:
\begin{multline}
k(\theta) = k_0 \Big(1 + \lambda_{\text{cod}} \cos(n \theta) \\
+\lambda_{\beta_x} \cos(p_x \theta) + \lambda_{\beta_y} \cos(p_y \theta) \Big),
\end{multline}
where $\lambda_{\text{cod}}$, $\lambda_{\beta_x}$, and $\lambda_{\beta_y}$ are the perturbation amplitudes for each harmonic.
According to Eqs.~(\ref{eq:convcond}) and (\ref{eq:beta_mod}), $n$ and $p_x/2$ should be close to $\nu_x$
to ensure efficient modulation with minimal perturbation.  
Similarly, $p_y/2$ should be close to $\nu_y$ for effective modulation of $\beta_y$.  
To avoid introducing strong nonlinearity and to preserve a wide dynamic aperture (see Sec.~\ref{sec:disc}),
we prefer to set $k_0$ relatively low, which means low $\nu_x$.
To independently modulate $\beta_x$ and $\beta_y$, we also require $\nu_y$ to differ from $\nu_x$ by at least an integer.
Therefore, we set the tunes as $\nu_x$=1.91 and $\nu_y$=3.08, as shown in Table~\ref{tab:main_parameter}.  
Consequently, the harmonic numbers are determined as $n$=2, $p_x$=4, and $p_y$=6.

\begin{figure*}[htb]
\centering
    \includegraphics[width=16cm]{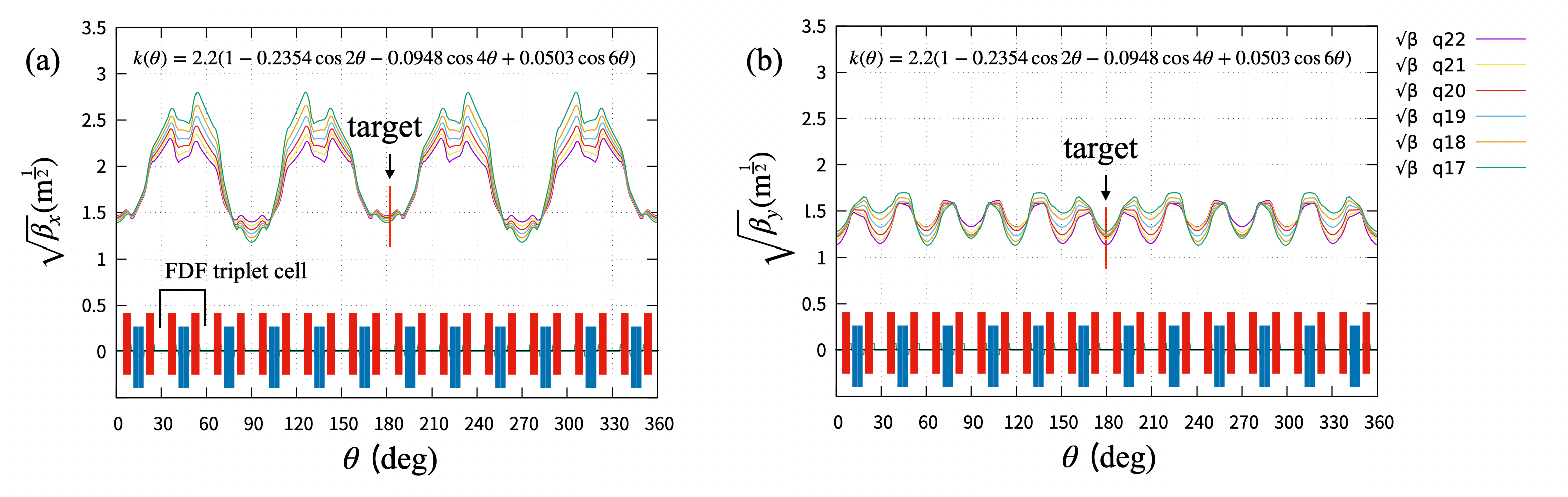}
    \caption{
        Betatron functions for charge states ranging from $q$=$17^+$ to $22^+$.  
        Panel (a) shows $\beta_x$, and panel (b) shows $\beta_y$, both obtained using the modulated~$k$.}
\label{fig:betas}
\end{figure*}

Next, we determined $\lambda_{\text{cod}}$, $\lambda_{\beta_x}$, and $\lambda_{\beta_y}$.  
Since these parameters are not mutually independent, they were numerically optimized to minimize the growth of horizontal and
vertical root mean square (RMS) emittances, $\epsilon_x$ and $\epsilon_y$, over 1,000 turns.
The optimization was carried out using the particle tracking code \textsc{EARLIETIMES}~\cite{suzie}, operated in
fourth-order Runge–Kutta mode under the hard-edge magnetic field approximation (i.e., fringe fields were neglected).  
In this simulation, 1,000 macro-particles were initialized with a betatron amplitude of $2J_x = 2J_y=\SI{100}{mm.mrad}$,
uniformly distributed along the Courant--Snyder ellipse in phase space.  
The corresponding RMS emittance of this distribution was \SI{50}{mm.mrad}.  
Each particle was initially assigned a charge state of $20^+$, and tracking began from $\theta = 0$, which corresponds to
\SI{180}{\degree} upstream from the target.  
At each target traversal, the particle's charge state was randomly re-assigned according to the probability
distribution shown in Fig.~\ref{fig:shima_dist}.  
Energy loss, multiple Coulomb scattering in the target, and synchrotron motion were all neglected in this optimization.
Space-charge effects were also ignored, as the resulting tune shift is negligible under the current beam conditions (see Sec.~\ref{sec:disc}).
The optimized perturbation amplitudes were found to be:  
$\lambda_{\text{cod}}$=$-0.2354$, $\lambda_{\beta_x}$=$-0.0948$, and $\lambda_{\beta_y}$=0.0503.

\subsection{Behavior of Closed Orbits, Betatron Functions, and emittance growths}
\label{subsec:charact_ring}

Figure~\ref{fig:orbits} shows the closed orbits for charge states ranging from $q = 17^+$ to $22^+$ under constant~$k$ (purple)
and modulated~$k$ (green) configurations.  
Due to the use of harmonic number $n$ = 2 for COD excitation, the modulated~$k$ ring generates two convergence/focus points around the ring,
effectively guiding particles toward stable orbits.
(Using a higher harmonic number (e.g., $n = 3$) would generate three convergence points, which could be advantageously used to
place RF cavities and would also reduce orbit excursions for different charge states.
However, achieving this requires the horizontal tune $\nu_x$ to be close to 3, which in turn demands a higher $k_0$,
introducing strong nonlinearities into the system.)

To evaluate the orbital stability, we examined the motion of a test particle initialized with zero betatron amplitude (i.e., $2J_x = 0$).  
Figure~\ref{fig:sherit}(a) shows the footprints of the test particle over 200 turns: the purple line represents
  the case with constant~$k$, while the green line corresponds to the modulated~$k$ configuration.  
Figure~\ref{fig:sherit}(b) shows the evolution of the betatron amplitude $2J_x$ for both cases.  
In the constant~$k$ ring, $2J_x$ grows rapidly, indicating strong instability.  
In contrast, the modulated~$k$ ring effectively suppresses the amplitude growth, and the particle remains oscillating in the vicinity
of the closed orbit, demonstrating significant improvement in orbit confinement.

Figure~\ref{fig:betas} presents the betatron functions $\beta_x$ and $\beta_y$ for charge states ranging from $q = 17^+$ to $22^+$.  
In both planes, the $\beta$-functions exhibit clear convergence at the target position, indicating successful optical matching across
multiple charge states.  
This result confirms that the modulated~$k$ configuration ensures robust focusing and charge-state-independent transport at the target.

\begin{figure}[htb]
\centering
    \includegraphics[width=8cm]{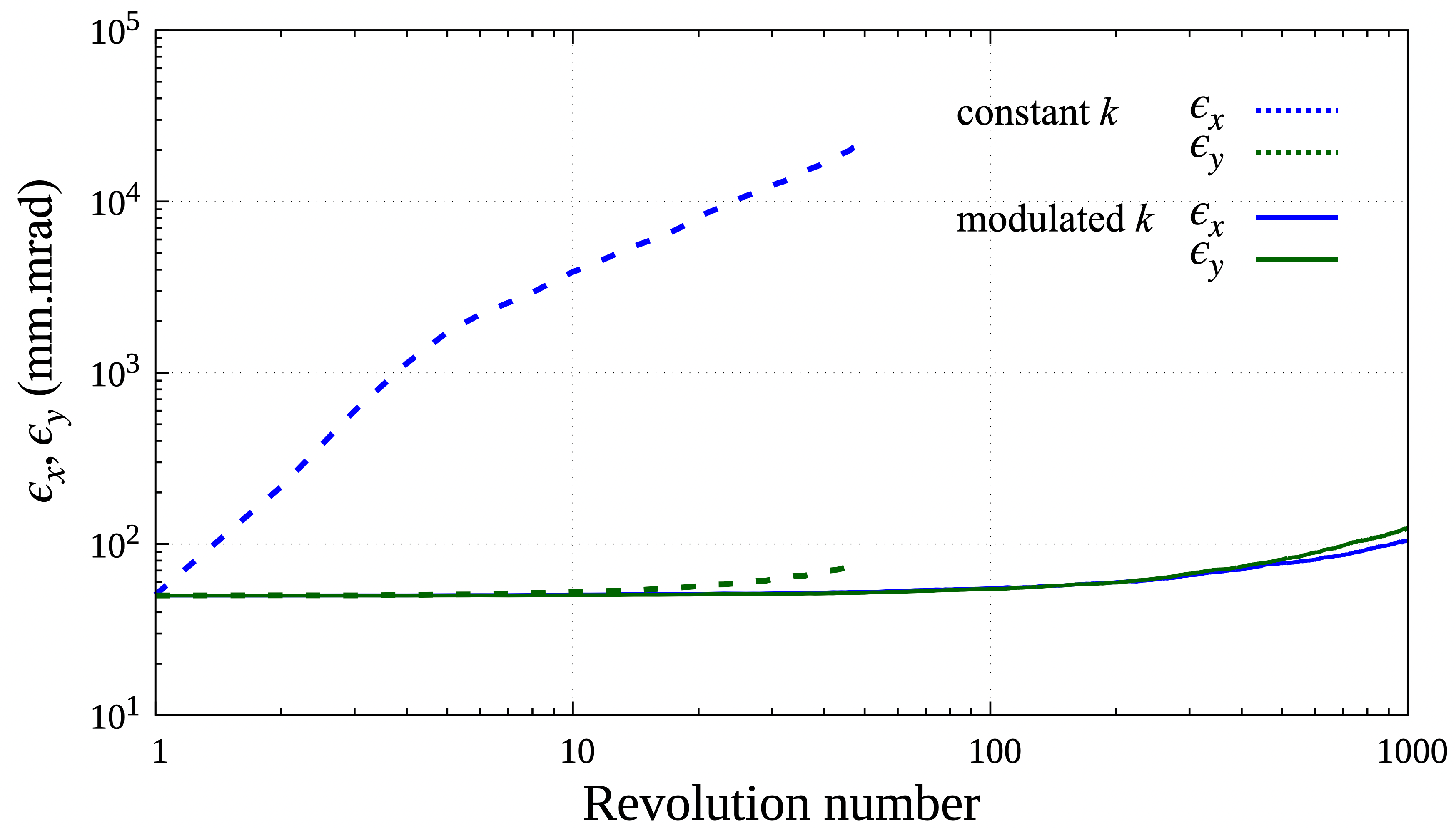}
    \caption{Emittance growth in the $x$ and $y$ planes, with (solid) and without (dashed) $k$ modulation.}
\label{fig:emit_gr}
\end{figure}

The effect of the modulation on beam stability is further illustrated in Fig.~\ref{fig:emit_gr}, which shows the evolution of $\epsilon_x$ (blue) and $\epsilon_y$ (green).
Solid lines indicate the case with $k$ modulation, while dashed lines represent the unmodulated case.  
Without $k$ modulation, $\epsilon_x$ increases rapidly, reaching approximately \SI{20000}{mm.mrad} within 50 turns.  
In contrast, with the optimized $k$ modulation, the emittance growth in both planes is significantly mitigated.  
After 1,000 turns, both $\epsilon_x$ and $\epsilon_y$ remain near \SI{100}{mm.mrad}—about only twice the initial value—clearly demonstrating
the effectiveness of the modulation in preserving beam quality.

%%%%%%%%%%%%%%%%%%%%%%%%%%
\section{Beam Simulation}
\label{sec:sim}
%%%%%%%%%%%%%%%%%%%%%%%%%%

\begin{figure}[htb]
\centering
\includegraphics[clip,width=8cm]{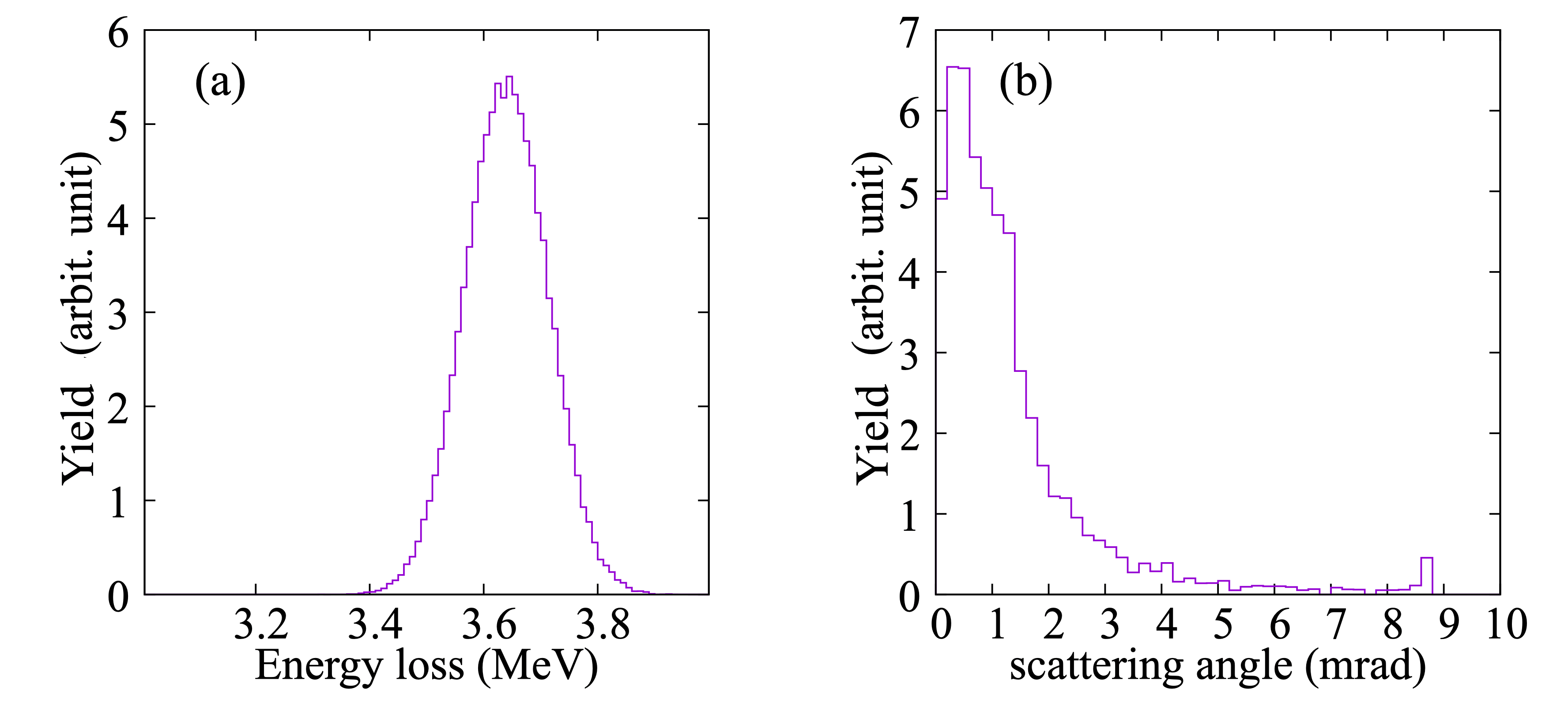}
\caption{Energy loss and scattering angle distributions used for the beam simulation, obtained by TRIM.}
\label{fig:e_loss_scat}
\end{figure}

Based on the lattice described above, a 6D beam-tracking simulation was performed.
The application under consideration is a storage ring for the production of superheavy elements.
We assumed that a \ce{^{50}Ti^{20+}} beam was incident on a \ce{^{249}Bk} target to produce element 119.
To account for beam–target interactions, the following effects were incorporated into the simulation:
(1) energy loss and straggling of the beam, and
(2) energy recovery by an rf cavity.
The SCSC process was switched on and off to evaluate its impact.
For energy loss and straggling, the Monte Carlo code TRIM~\cite{TRIM} was used.
Figure~\ref{fig:e_loss_scat}(a) shows the energy loss distribution of the \ce{^{50}Ti} beam (\SI{5.5}{MeV/u}) after passing through the target at the incident beam energy of \SI{275}{MeV}, while Fig.~\ref{fig:e_loss_scat}(b) shows the corresponding scattering angle distribution.
The simulation parameters are summarized in Table~\ref{tab:sim_param}.
\begin{table}[htb]
\setlength\tabcolsep{5.5pt}
\centering
\caption{Parameters used in the beam simulation.}
\label{tab:sim_param}
\begin{tabular}{lc}
\hline
\hline
Beam species & {\ce{^{50}Ti^{q+}} (q = 17--22)} \\
Beam energy & \SI{275}{MeV} \\
RF cavity voltage & \SI{1.2}{MV} \\
RF cavity frequency & \SI{0.861}{MHz} \\
Harmonic number & 1 \\
Synchronous phase & \SI{8.7}{\degree} \\
Slippage factor & $-0.676$ \\
Target material & \ce{^{249}Bk} \\
Target thickness & \SI{100}{\micro g/cm^2} \\
Injection beam duty factor & 20\% \\
Number of macro particles & 1000 \\
Physical aperture & $\SI{4.5}{m} < r < \SI{7.5}{m}$ \\
& $\SI{-0.15}{m} < y < \SI{0.15}{m}$ \\
\hline
\hline
\end{tabular}
\end{table}
The rf cavity was located at $\theta$ = \SI{0}{\degree}.
\begin{figure*}[htb]
\centering
\includegraphics[clip,width=16cm]{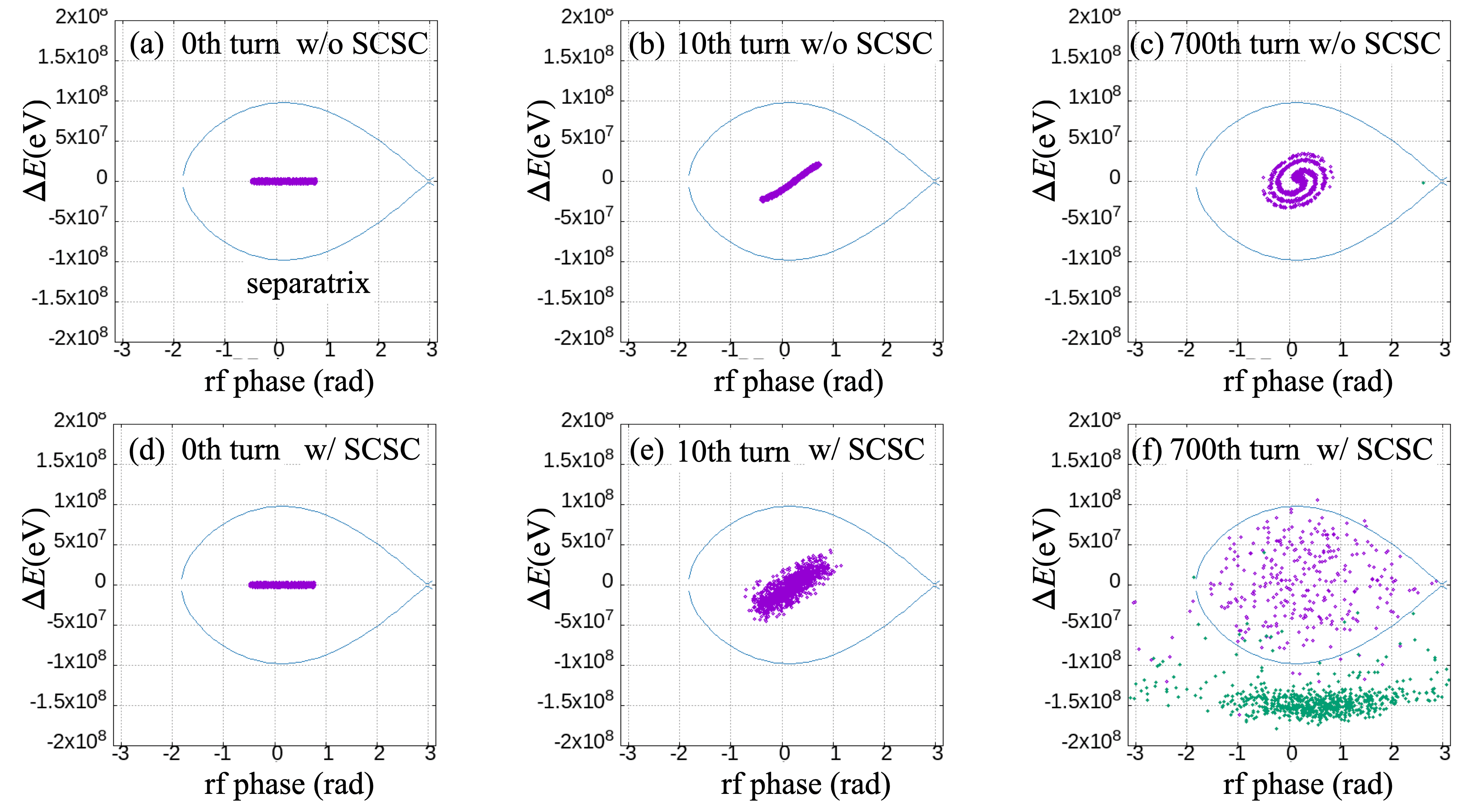}
\caption{Time evolution of the longitudinal phase space ($\Delta E$--$\phi_{\text{rf}}$).
Panels (a)–(c) show snapshots at the 0th, 10th, and 700th turns without SCSC effects.
Panels (d)–(f) show corresponding snapshots with SCSC effects included.
Green dots indicate particles that exceeded the radial aperture limit.}
\label{fig:fish}
\end{figure*}

\begin{figure}[htb]
\centering
\includegraphics[clip,width=8cm]{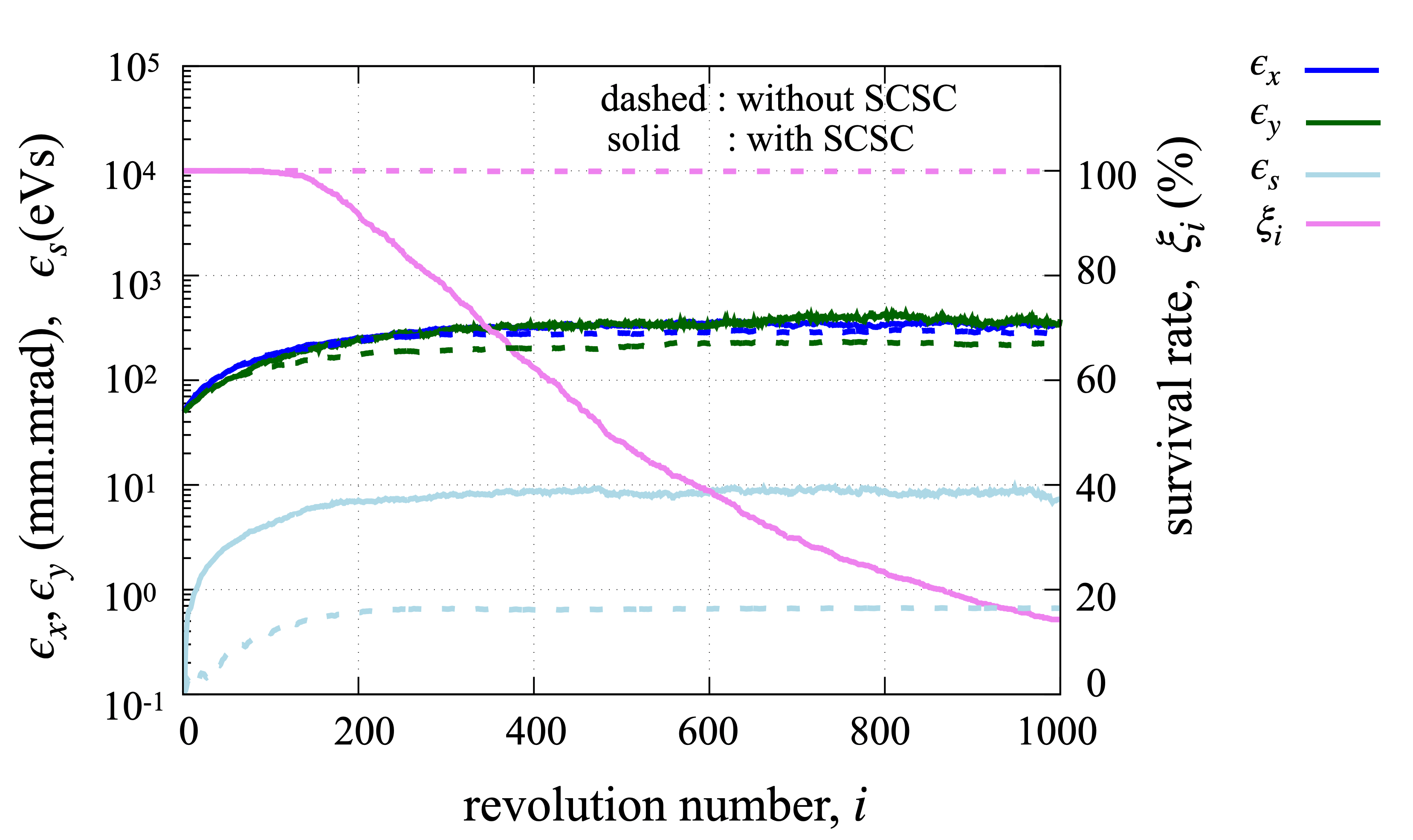}
\caption{Emittance growth and survival rate. Blue, green, and light blue lines represent $\epsilon_x$,
  $\epsilon_y$, and $\epsilon_s$, respectively.
  Magenta lines represent the survival rate~$\xi_i$.
  Dashed and solid lines correspond to cases without and with SCSC effects, respectively.}
\label{fig:emit_grs_6d}
\end{figure}

The simulation results are presented in Figs.~\ref{fig:fish} and \ref{fig:emit_grs_6d}.
Figure\ref{fig:fish} shows the time evolution of the longitudinal phase space ($\Delta E$--$\phi_{\text{rf}}$).
The upper and lower rows correspond to the cases without and with SCSC effects, respectively.
Due to energy loss at the target, the synchronous phase was set to \SI{8.7}{\degree}, as listed in Table~\ref{tab:sim_param}, and an acceleration bucket was formed accordingly.
The separatrix shown corresponds to the $q=20^+$ charge state.
As shown in panels (a) and (d), the beam was injected with a 20\% duty factor.
A comparison between (b) and (e) clearly shows that SCSC acts as a diffusive process in longitudinal phase space, driving particles beyond the separatrix.
Once particles leave the stable region, the ERIT mechanism can no longer compensate for their energy loss at the target.
Consequently, these particles continue to lose energy and their orbits shift inward.
Eventually, they reach the radial aperture limit, as indicated by the green dots in panel (f).

Figure~\ref{fig:emit_grs_6d} shows the evolution of the RMS emittance in the transverse($\epsilon_x$,
$\epsilon_y$), and longitudinal ($\epsilon_s$) directions, as well as the survival rate~$\xi_i$ at each turn $i$.
Solid and dashed lines represent cases with and without SCSC, respectively.

In the absence of SCSC, $\epsilon_s$ increased during the first $\sim$200 turns due to filamentation in phase space, and then saturated below \SI{1}{eVs}.
Due to the scattering effects, $\epsilon_x$ and $\epsilon_y$ also increased up to several hundred \SI{}{mm.mrad} and then saturated.
Their growth was mitigated by ionization cooling~\cite{ERIT_okabe}, resulting in no beam loss.

When SCSC was included, $\epsilon_s$ increased rapidly due to diffusion in the longitudinal phase space (as shown in Fig.~\ref{fig:fish}).
The survival rate $\xi_i$ began to decrease after approximately 100 turns, indicating beam loss caused by this diffusion process.
Nevertheless, it is noteworthy that about 15\% of the particles still survived after 1000 turns.

\begin{figure}[htb]
\centering
\includegraphics[clip,width=8cm]{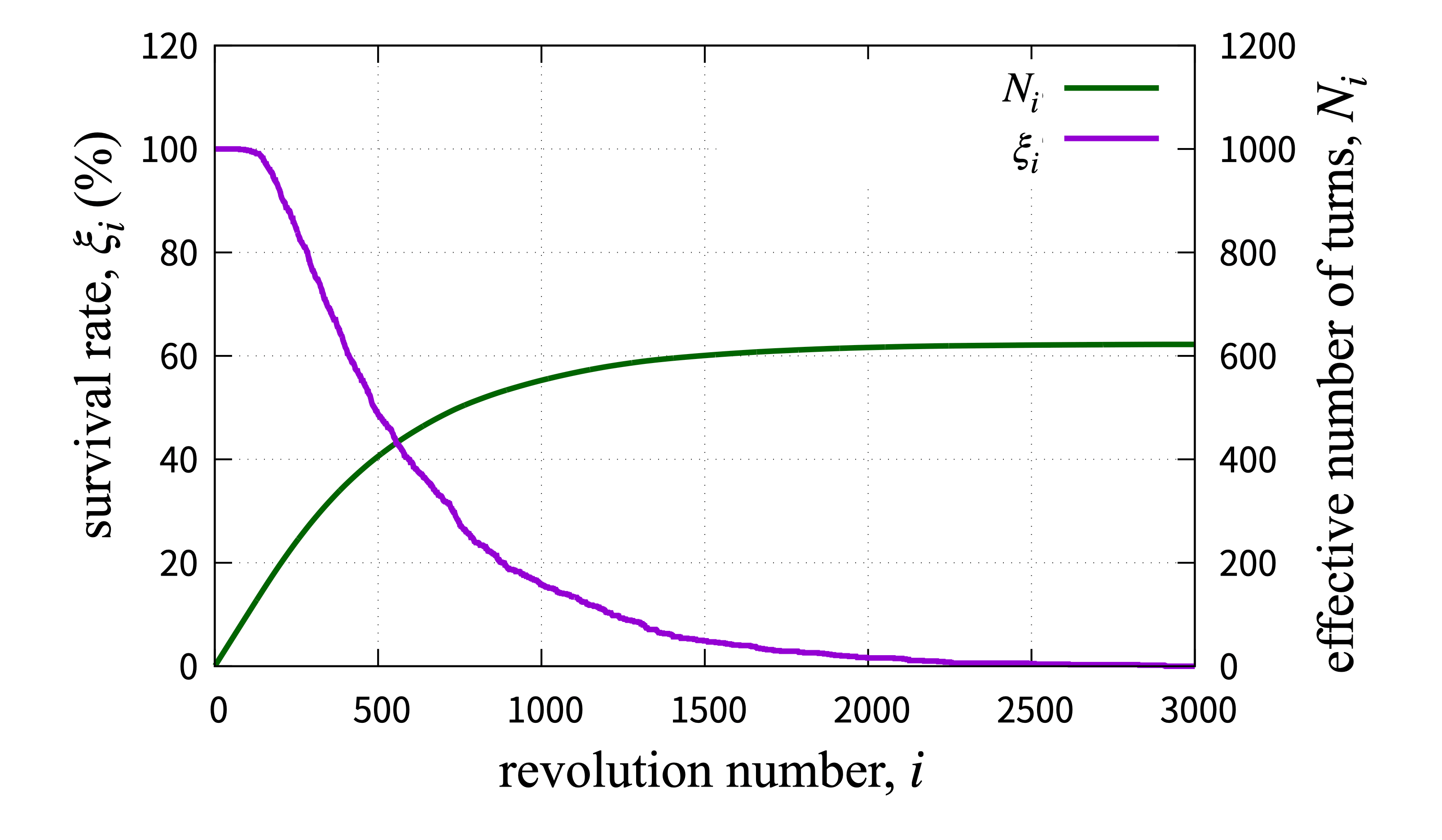}
\caption{Survival rate $\xi_i$ and effective number of turns $N_i$.}
\label{fig:eff_turn}
\end{figure}

In typical ERIT operation, a continuous beam is injected from an upstream accelerator into the storage ring to irradiate the internal target.
Accordingly, the effective number of turns $N_i$ is defined as the weighted sum of the surviving beam at each turn:
\begin{align}
N_i = \sum_{j}^{i} j~\xi_j.
\end{align}
Figure~\ref{fig:eff_turn} shows $\xi_i$ (purple) and the accumulated effective turns $N_i$ (green).
The results show that $N_i$ saturates at approximately 600 around the 1500th turn.
%%%%%%%%%%%%%%%%%%%%%%%%%%%%%%%%%%%%%%%%%%%%%%%%%%%%
\section{discussion}
\label{sec:disc}
%%%%%%%%%%%%%%%%%%%%%%%%%%%%%%%%%%%%%%%%%%%%%%%%%%%%
In this section, we discuss the chromaticity and transverse beam stability associated with the use of modulated~$k$.
We also address the space-charge effects, which are unavoidable in low-energy storage rings.

%-----------------------------
\subsection{Chromaticity and transverse beam stability}
%-----------------------------
An ideal scaling FFA ring exhibits zero chromaticity, ensuring identical tunes for beams with different charge states. However, when the scaling condition is violated—such as through the modulation of~$k$—this chromaticity is no longer guaranteed. It is therefore essential to evaluate how the tune varies with charge state under such conditions.

\begin{figure}[htb]
\centering
\includegraphics[width=7cm]{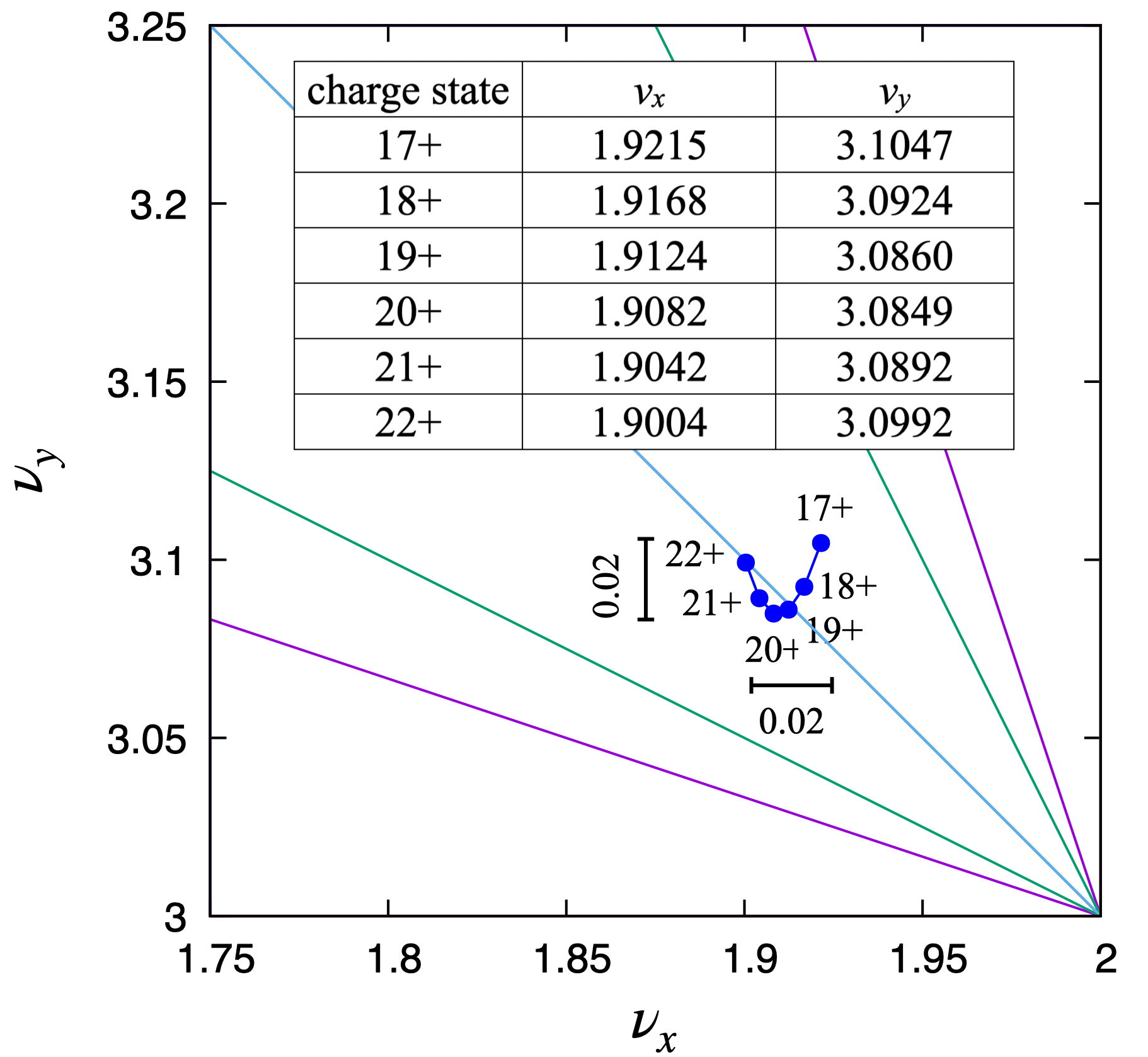}
\caption{Tune variation for beams with different charge states plotted on the ring’s tune diagram. The diagram includes resonance lines up to fourth order for a ring with superperiodicity 1.}
\label{fig:tune}
\end{figure}

Figure~\ref{fig:tune} shows the horizontal and vertical tunes ($\nu_x$ and $\nu_y$) for six charge states ranging from 17$^+$ to 22$^+$ (see also Table~\ref{tab:main_parameter} for the reference charge state $q_0$). The resulting tune variation is approximately 0.02, which is sufficiently small to conclude that the deviation from ideal zero chromaticity poses no significant issue.

Integer resonances are employed for closed-orbit matching, while half-integer resonances are utilized for beta function matching. The tunes are chosen close to these resonance harmonics so that modulations of the closed orbit and beta functions are enhanced with minimal perturbation. Under this condition, the magnetic field distribution effectively excites the relevant resonances. Nonetheless, as demonstrated in Figs.~\ref{fig:cod_matching} and \ref{fig:betas}, these excitations remain sufficiently controlled: there is no instability in the closed orbit due to integer resonance, nor divergence in amplitude from half-integer resonance excitation.

The introduction of $k$ modulation breaks the ring symmetry, reducing it from 12-fold to 2-fold symmetry in the test ring. This prompts an important question: to what extent does the reduced symmetry and associated resonance excitation affect transverse phase-space stability?

\begin{figure}[htb]
\centering
\includegraphics[clip,width=8cm]{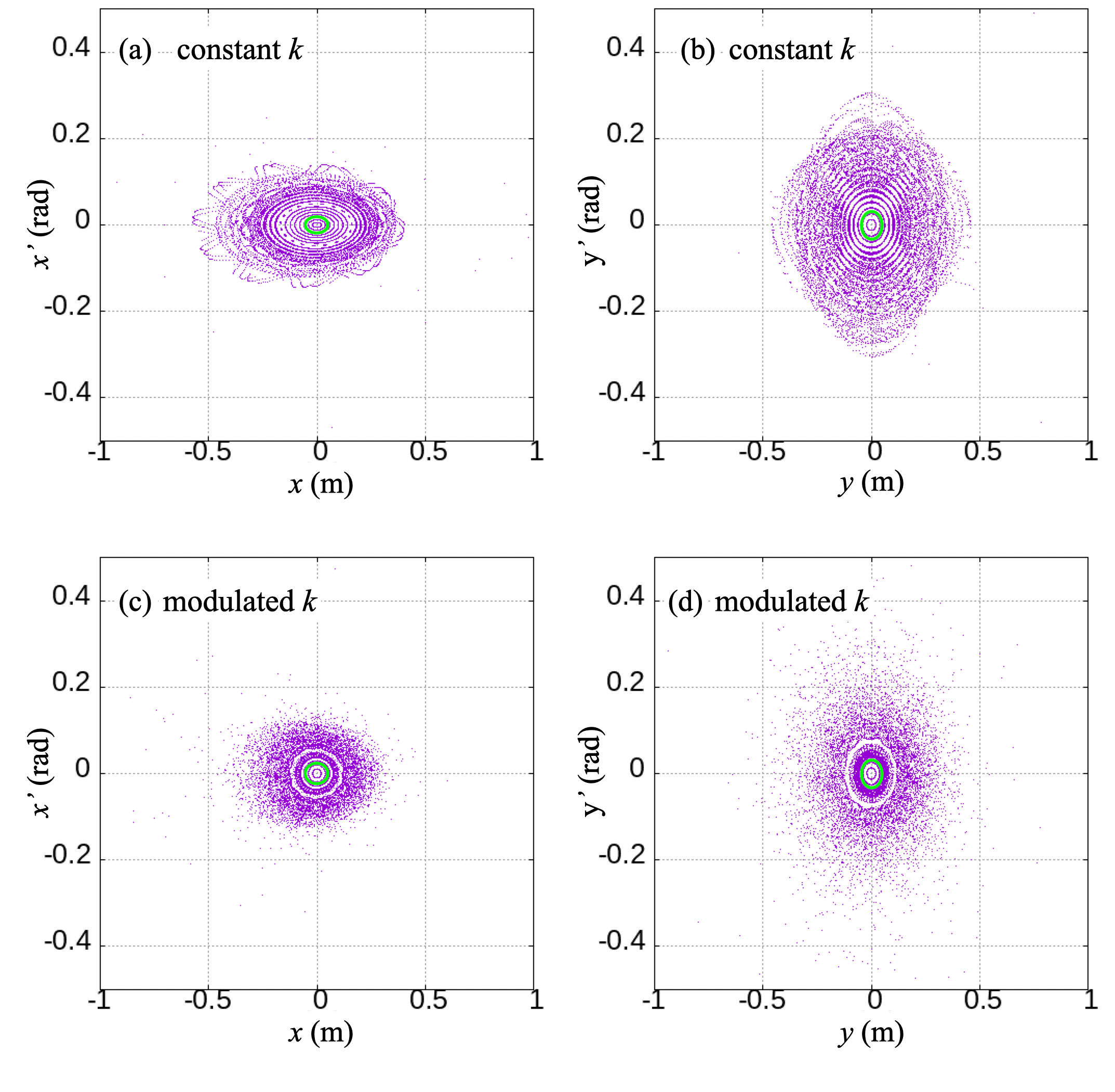}
\caption{Transverse phase-space structure at the target location without SCSC effects. Upper: constant~$k$. Lower: modulated~$k$.The green ellipses correspond to
 the apertures, $x$=$\pm$\SI{50}{mm}, $y$=$\pm$\SI{50}{mm}.}
\label{fig:dynap}
\end{figure}

Figure~\ref{fig:dynap} compares the phase-space structures before and after $k$ modulation.
The green ellipses indicate the apertures  $x$=$\pm$\SI{50}{mm}, $y$=$\pm$\SI{50}{mm}.
Despite the lowered symmetry, a sufficiently large stable region is maintained in both horizontal and vertical directions,
indicating that transverse stability is preserved under modulation.

%-----------------------------
\subsection{Space charge effects}
%-----------------------------
The influence of space-charge on the tune is estimated using the following expressions:
\begin{align}
\Delta \nu_x &= -\frac{r_InR}{\nu_x\pi\beta^2\gamma^3}\frac{1}{\sigma_x(\sigma_x+\sigma_y)B_f}, \notag \\
\Delta \nu_y &= -\frac{r_InR}{\nu_y\pi\beta^2\gamma^3}\frac{1}{\sigma_y(\sigma_x+\sigma_y)B_f}, \\
\sigma_x   &= \sqrt{\tilde{\beta_x}\epsilon_x+[D_x(\Delta p/p_0)]^2} ,\notag  \\
\sigma_y   &= \sqrt{\tilde{\beta_y}\epsilon_y} \notag ,
\end{align}
where $r_I$ is the classical radius of the ion, $n$ is the total number of stored particles,
$R$ is the average radius of the ring, and $B_f$ is the bunching factor. The transverse beam sizes $\sigma_x$ and
$\sigma_y$ include contributions from both emittance and momentum dispersion.
A \ce{^{50}Ti^{20+}} beam with an energy of \SI{275}{MeV} and a current of \SI{1}{p\micro A} was assumed in the evaluation.
The average beta functions were set to $\tilde{\beta}_x = \SI{3.2}{m}$ and $\tilde{\beta}_y = \SI{1.9}{m}$,
with transverse emittances of \SI{400}{mm.mrad} in both planes. The bunching factor was assumed to be $B_f = 0.2$.
Allowing for a maximum space-charge tune shift of 0.1, the number of stored particles could reach up to $1.5 \times 10^{10}$.
However, in the test ring, beam losses limited the effective number of turns to about 600, well below the space-charge limit.
Thus, beam accumulation was constrained more by particle loss than by space-charge effects.

%%%%%%%%%%%%%%%%%%%%%%%%%%%%%%%%%%%%%%%%%%%%%%%%%%%%
\section{conclusion}
\label{sec:concl}
%%%%%%%%%%%%%%%%%%%%%%%%%%%%%%%%%%%%%%%%%%%%%%%%%%%%

The ERIT-type heavy-ion storage ring is a promising device for efficient production of secondary particles in rare nuclear reactions.
One of the major challenges for stable ring operation is the SCSC that occurs when heavy-ion
beams interact with the internal target. This process induces random changes in charge states, which in turn cause significant
emittance growth and beam loss.

To address this issue, we developed a method for closed-orbit matching at the target location that accommodates multiple charge
states by utilizing coherent CODs. This approach is implemented within a scaling FFA optics framework.

Although the field index $k$ is globally constant in a scaling FFA, introducing a modulation of $k$ enables
orbit convergence at the target. A similar modulation strategy was applied to match the beta function. We confirmed that
beta-function matching is feasible by taking advantage of coherent beta-beat resonances induced by the modulated~$k$.
In simulations using a validation lattice, emittance growth due to SCSC was effectively suppressed: the initial transverse emittance
of $\epsilon_{x/y} = \SI{50}{mm.mrad}$ increased only to $\sim \SI{100}{mm.mrad}$ after 1000 turns.

Based on this lattice, we conducted full 6D beam tracking simulations.
The results demonstrated that the system can accumulate beam equivalent to over 600 effective turns.

The chromaticity in this system remains extremely small due to the scaling nature of the FFA optics. Even for a broad
charge state distribution ranging from $q=17^+$ to $22^+$—corresponding to a 25\% momentum deviation—the tune variation was
limited to only 0.02.% (see Fig.~\ref{fig:tune}).
In addition, we confirmed that the wide dynamic aperture, a key advantage of
scaling FFAs, was preserved even with the modulated $k$.

When applied to superheavy element production (e.g., elements 119 or 120), the space-charge tune shift remains below 0.1,
allowing for up to 2000 turns of accumulation. Since the effective number of turns achieved in this study was 600,
beam loss--not space charge--was the dominant limiting factor.

In contrast to the transverse direction, no comprehensive solution has yet been found for mitigating longitudinal beam
diffusion due to SCSC. Future work will explore potential remedies such as second-harmonic rf systems, barrier buckets,
and negative-$k$ lattices to suppress longitudinal diffusion.

In summary, our results demonstrate that scaling FFA rings with modulated~$k$ offer a viable and robust approach for
heavy-ion storage rings where SCSC is unavoidable.

\bibliography{sheerit}% Produces the bibliography via BibTeX.

\end{document}